\documentclass{appolb}

\usepackage{amssymb}
\usepackage{amsmath}
\usepackage{epsfig}
\usepackage{amsfonts}
\usepackage{graphicx}%
\begin{document}
\title{Quarkonium and glueball admixtures of the scalar-isoscalar resonances $f_{0}(1370), f_{0}(1500)$ and $f_{0}(1710)$%
\thanks{Presented at the Workshop \textquotedblleft Excited QCD 2013\textquotedblright,
Sarajevo/Bosnia and Herzegovina, February 3-9, 2013 }%
}
\author{Stanislaus Janowski
\address{Institute for Theoretical Physics, Goethe-University,
Max-von-Laue-Str.\ 1, D--60438 Frankfurt am Main, Germany}
%\and
%the Name(s) of other Author(s)
%\address{and their affiliation}
}
\maketitle
\begin{abstract}
Using the $U(3)_{R}\times U(3)_{L}$ extended Linear Sigma Model with
the ordinary (pseudo)scalar and (axial)vector mesons as well as a scalar glueball,
we study the vacuum phenomenology of the scalar-isoscalar resonances
$f_{0}(1370), f_{0}(1500)$ and $f_{0}(1710)$.
We present here a solution, based only on the masses and not yet on decays, in which the resonances
$f_{0}(1370)$ and $f_{0}(1500)$ are predominantly nonstrange and strange $\bar{q}q$ states
respectively, and the resonance $f_{0}(1710)$ is predominantly a scalar glueball. 
\end{abstract}

\section{Introduction}

The experimental verification of many scalar-isoscalar resonances,
\linebreak $I^{G}(J^{PC})=0^{+}(0^{++})$,
up to the energy of $2$ GeV \cite{PDG} reenforced the idea that the scalar glueball lies among them.
However, the remaining open question is which of these
scalars-isoscalars is predominantly the glueball.\newline
In this paper we address this issue by studying the vacuum phenomenology of scalar-isoscalar
states within the framework of the so called extended Linear Sigma Model (eLSM) \cite{dick, dilaton, susden}.
Our numerical calculations of the effective Lagrangian (\ref{Lagrangian})
are explicitly done in the case of three quark flavours.
For $N_{f}=3$ two scalar-isoscalar $\bar{q}q$ states exist, the nonstrange
$\sigma_{N}\equiv\left(\bar{u}u+\bar{d}d\right) /\sqrt{2}$ and the strange
$\sigma_{S}\equiv\bar{s}s$ mesons. The scalar glueball $G \equiv gg$
is implemented in the eLSM from the very beginning as the fluctuation of the dilaton field.
We describe $G$ as well as the generation of the scale anomaly of the pure Yang-Mills Lagrangian
at the quantum level by the usual logarithmic dilaton Lagrangian,
first introduced in Ref. \cite{schechter, migdal}. Due to the same quantum numbers a mixing
between the scalar-isoscalar quarkonia and glueball in our model takes place.\newline
In this study we focus on the determination of the mixing matrix by using the mass eigenvalues of the
scalar-isoscalar states and identify them with the physical masses of the
resonances $f_{0}(1370)$, $f_{0}(1500)$ and $f_{0}(1710)$ \cite{PDG}.\newline
The outline of this proceeding is as follows: In Sec. 2 we present the Lagrangian of the extended
Linear Sigma Model with a scalar glueball. In Sec. 3 we discuss our results and finally
in Sec. 4 we give a summary as well as an outlook for further research.

\section{The effective Lagrangian}

In order to study the vacuum phenomenology of the scalar-isoscalar resonances
$f_{0}(1370),f_{0}(1500)$ and $f_{0}(1710)$
we use the eLSM including a scalar glueball \cite{dick, dilaton, susden}.
Its compact representation for an optional number of flavours is as follows:

\begin{align}%
%TCIMACRO{\tciLaplace}%
%BeginExpansion
\mathcal{L}%
%EndExpansion
&  =\mathcal{L}_{dil}+\text{\textrm{Tr}}\left[  (D^{\mu}\Phi)^{\dag}(D_{\mu
}\Phi)-m_{0}^{2}\left(  \frac{G}{G_{0}}\right)  ^{2}\Phi^{\dag}\Phi
-\lambda_{2}(\Phi^{\dag}\Phi)^{2}\right]\nonumber\\ & -\lambda_{1}(\text{\textrm{Tr}}%
[\Phi^{\dag}\Phi])^{2}+c_{1}(\text{\textrm{det}}\Phi^{\dag}-\text{\textrm{det}}
\Phi)^{2}+\mathrm{Tr}[H(\Phi^{\dag}+\Phi)]\nonumber\\
&  +\text{\textrm{Tr}}\left[  \left(  \frac{m_{1}^{2}}{2}\left(  \frac
{G}{G_{0}}\right)  ^{2}+\Delta\right)  \left(L^{\mu2}+R^{\mu2}\right)  \right] \nonumber\\
& -\frac{1}{4}\text{\textrm{Tr}}\left[
L^{\mu\nu2}+R^{\mu\nu2}\right]  +\frac{h_{1}}{2}\text{\textrm{Tr}%
}[\Phi^{\dag}\Phi]\text{\textrm{Tr}}[L_{\mu}L^{\mu}+R_{\mu}R^{\mu}]\nonumber\\
&  +h_{2}\text{\textrm{Tr}}[\Phi^{\dag}L_{\mu}L^{\mu}\Phi+\Phi R_{\mu}R^{\mu
}\Phi^{\dag}]+2h_{3}\text{\textrm{Tr}}[\Phi R_{\mu}\Phi^{\dag}L^{\mu}]\text{
+... ,}\label{Lagrangian}
\end{align}
\newline
where
\begin{equation}
D^{\mu}\Phi=\partial^{\mu}\Phi-ig_{1}(L^{\mu}\Phi-\Phi
R^{\mu})
\end{equation}
and
\begin{equation}
\mathcal{L}_{dil}=\frac{1}{2}(\partial_{\mu}G)^{2}-\frac{1}{4}\frac{m_{G}^{2}%
}{\Lambda^{2}}\left(  G^{4}\ln\left\vert \frac{G}{\Lambda}\right\vert
-\frac{G^{4}}{4}\right)  \text{.}\label{ldil}%
\end{equation}\newline
The dilaton Lagrangian (\ref{ldil}) describes the scale anomaly
of the pure Yang-Mills sector where the logarithmic term with the energy scale $\Lambda$ breaks
the dilatation symmetry explicitly. The field $G$ is the scalar dilaton and after performing the shift
$G\rightarrow G_{0}+G$ a particle with the mass $m_G$ arises which we interpret as the scalar glueball.\newline
Our model underlies the following symmetries and their several breakings:
i) The global chiral symmetry, $U(N_{f})_{R}\times U(N_{f})_{L}$, which is broken explicitly due to
the quantum effects and the nonvanishing quark masses as well as spontaneously.
As a consequence of breaking of the latter one a nonvanishing quark condensate, $\left\langle\bar{q}q\right\rangle\neq0$, arises. \newline
ii) A crucial symmetry in our model is the already mentioned dilatation symmetry or scale invariance, 
$x^{\mu}\rightarrow\lambda^{-1}x^{\mu}$, which is realized 
at the classical level of the Yang-Mills sector of QCD, but explicitly broken by the loop corrections.
This is known as scale or trace anomaly, respectively and leads to the nonvanishing gluon
condensate, $\left\langle \frac{\alpha_{s}}{\pi}\,G_{\mu\nu}^{a}G^{a,\mu\nu}\right\rangle\neq0$.
Taking into account the dilatation symmetry we can constrain the number of possible terms in our model.
This implies that in the chiral limit with the exception of the logarithmic term in Eq. (\ref{ldil})
and the term generating the $U(1)_{A}$ anomaly, all parameters entering in the model are dimensionless.
(Note, using the chiral anomaly it was also possible to couple the pseudoscalar glueball to the model, for details and
results see Refs.\cite{psg}).\newline
iii) Our model is also in agreement with discrete symmetries of QCD e.g. parity $P$ and charge conjugation $C$.\newline
The multiplet of the ordinary scalar and pseudoscalar mesons in the case of $N_{f}=3$, containing the two out of three
bare scalar-isoscalar states $\sigma_{N}$ and $\sigma_{S}$, reads \cite{dick}:

\begin{equation}
\Phi=\frac{1}{\sqrt{2}}\left(
\begin{array}
[c]{ccc}%
\frac{(\sigma_{N}+a_{0}^{0})+i(\eta_{N}+\pi^{0})}{\sqrt{2}} & a_{0}^{+}%
+i\pi^{+} & K_{0}^{\star+}+iK^{+}\\
a_{0}^{-}+i\pi^{-} & \frac{(\sigma_{N}-a_{0}^{0})+i(\eta_{N}-\pi^{0})}%
{\sqrt{2}} & K_{0}^{\star0}+iK^{0}\\
K_{0}^{\star-}+iK^{-} & \bar{K}_{0}^{\star0}+i\bar{K}^{0} & \sigma_{S}+i\eta_{S}%
\end{array}
\right). \label{phimatex}%
\end{equation}
\newline
The explicit form of the left-handed and the right-handed (axial)vector mesons multiplets, $L^{\mu}$ and $R^{\mu}$ 
can be found in Ref. \cite{dick}.

\section{Results and Discussion}

A reasonable approach to study the scalar-isoscalar sector of the eLSM with a scalar glueball
is to use the values of the global fit performed in Ref. \cite{dick} and only to determine
the additional three free parameters which enter in this sector.
Two of them arise directly from the dilaton Lagrangian (\ref{ldil}), namely the bare mass of the scalar glueball
$m_{G}$ and the energy scale $\Lambda$. The third one is $\lambda_{1}$ which couples the ordinary (pseudo)scalar mesons
and was in Ref. \cite{dick} only determined as a part of combination.\newline
In order to obtain the numerical values of the corresponding parameters we use the bare masses of the scalar-isoscalar states
and the experimental masses of the resonances $f_{0}(1370), f_{0}(1500)$ and $f_{0}(1710)$.
Accordingly we consider the potential of the corresponding states in the matrix representation:

\begin{equation}
V(\sigma_{N},G,\sigma_{S})=\frac{1}{2}(\sigma_{N},G,\sigma_{S})\left(
\begin{array}
[c]{ccc}%
m_{\sigma_{N}}^{2} & z_{G\sigma_{N}} & z_{\sigma_{S}\sigma_{N}}\\
z_{G\sigma_{N}} & M_{G}^{2} & z_{G\sigma_{S}}\\
z_{\sigma_{S}\sigma_{N}} & z_{G\sigma_{S}} & m_{\sigma_{S}}^{2}%
\end{array}
\right) \left(
\begin{array}
[c]{c}%
\sigma_{N}\\
G\\
\sigma_{S}%
\end{array}
\right)\label{pot}.%
\end{equation}
\newline
The corresponding bare mass equations are:

\begin{equation}
m_{\sigma_{N}}^{2}=C_{1}+2\lambda_{1}\phi_{N}^{2}+\frac{3}{2}\lambda_{2}\phi_{N}^{2} \text{ ,}\label{msn}
\end{equation}
\begin{equation}
M_{G}^{2}=\frac{m_{0}^{2}}{G_{0}^{2}}(\phi_{N}^{2}+\phi_{S}^{2})+\frac
{m_{G}^{2}}{\Lambda^{2}}\left(  1+3\ln\left\vert \frac{G_{0}}{\Lambda
}\right\vert \right)  G_{0}^{2} \text{ ,}\label{Mg}
\end{equation}
\begin{equation}
m_{\sigma_{S}}^{2}=C_{1}+2\lambda_{1}\phi_{S}^{2}+3\lambda_{2}\phi_{S}^{2} \text{ .}\label{mss}%
\end{equation}\newline 
After diagonalization of the bare mass matrix (see Eq. (\ref{pot})) we obtain the matrix containing the physical masses
of the scalar-isoscalar states, $M'=BMB^{T}$, where $B$ is the mixing matrix which transforms
the bare states into the physical ones and vice versa:

\begin{equation}
\left(
\begin{array}
[c]{c}%
f_{0}(1370)\\
f_{0}(1710)\\
f_{0}(1500)%
\end{array}
\right)=B
\left(
\begin{array}
[c]{c}%
\sigma_{N}\equiv\left(\bar{u}u+\bar{d}d\right) /\sqrt{2}\\
G\equiv gg\\
\sigma_{S}\equiv\bar{s}s\\
\end{array}
\right)  \text{ .}\label{trafo}%
\end{equation}
The numerical values of the free parameters are presented in Table \ref{Table1}.
They have been obtained by requiring that the three following masses of the scalar-isoscalar states \cite{PDG} hold:
$M_{f_{0}(1370)}=(1350\pm150)$ MeV, $M_{f_{0}(1500)}=(1505\pm6)$ MeV and $M_{f_{0}(1710)}=(1720\pm6)$ MeV.

\begin{table}[h] \centering
%EndExpansion%
\begin{tabular}
[c]{|c|c|c|}\hline
Parameter & Value \\\hline
$\lambda_{1}$ & $2.03$ \\\hline
$m_{G}$ & $1580$ [MeV] \\\hline
$\Lambda$ & $930$ [MeV]\\\hline
\end{tabular}
\caption{Values of the free parameters.\label{Table1}}
\end{table}
\noindent
The numerical solution of the mixing matrix reads:

\begin{equation}
B=\left(
\begin{array}
[c]{ccc}%
0.92 & -0.39 & 0.05\\
-0.33 & -0.83 & -0.45\\
-0.22 & -0.40 & 0.89%
\end{array}
\right) \text{ .}\label{bnum}
\end{equation}\newline
It turns out that the resonances
$f_{0}(1370)$ and $f_{0}(1500)$ are predominantly nonstrange and strange $\bar{q}q$ states,
and that the resonance $f_{0}(1710)$ is predominantly a scalar glueball.
For other results of the mixing matrix $B$ using different theoretical models, we refer to Refs. \cite{glueballs} and refs. therein.
The solution in Eq. \ref{bnum} relies only on calculations of masses of the scalar-isoscalar states.
In order to give final statements regarding the composition of the $f_{0}$ resonances evaluation of the corresponding 
decays is essential. Nevertheless, it is astonishing that by increasing the number of flavours in our model,
$N_{f}=2 \rightarrow N_{f}=3$, the phenomenology of the $f_{0}$ resonances seems to change (see Ref. \cite{dilaton}).

\section{Conclusions and Outlook}

We have evaluated the numerical values of the mixing matrix $B$ by using 
the masses of the scalar-isoscalar states between $1$ and $2$ GeV
within a chiral model, the so called extended Linear Sigma Model with a scalar glueball \cite{dick, dilaton}. We have found 
that the resonances $f_{0}(1370)$ and $f_{0}(1500)$ are predominantly nonstrange and strange $\bar{q}q$ states,
and the resonance $f_{0}(1710)$ is predominantly a scalar glueball.\newline
Study in progress are calculations of the decay processes appropriate to the solution presented in this work and 
further search for a solution with the assignment where $f_{0}(1500)$ is predominantly a glueball and $f_{0}(1710)$ predominantly
a strange $\bar{q}q$ state \cite{dünn}.

\section*{Acknowledgments}

The author thanks F. Giacosa, D. Parganlija and D. H. Rischke for cooperation and
useful discussions and acknowledges support from H-QM and HGS-HIRe.

\end{document}